\documentstyle[aps,prb,epsf,amsmath]{revtex} 

\begin{document}
\draft

\title{Effect of Substitutional Impurities on the Electronic States
  and Conductivity of Crystals with Half--filled Band}
\author{E. P.  Nakhmedov$^{1,2}$, H. Feldmann$^1$, R.  Oppermann$^1$ and M. Kumru$^3$}
\address{$^1$Institut f\"ur Theoretische Physik, Universit\"at W\"urzburg,
  97074 W\"urzburg, F.R.Germany\\
$^2$Azerbaijan Academy of Sciences, Institute of Physics, H. Cavid St. 33, Baku, Azerbaijan\\
$^3$ Fatih University, B\"uy\"ukcekmece, Istanbul, Turkey} \date{\today} \maketitle

\begin{abstract}
  Low temperature quantum corrections to the density of states (DOS) and the conductivity are examined for a two--dimensional(2D) square crystal with substitutional impurities. By summing the leading logarithmic corrections to the DOS its energy dependence near half--filling is obtained. It is shown that substitutional impurities do not suppress the van Hove singularity at the middle of the band, however they change its energy dependence strongly.

Weak disorder due to substitutional impurities in the three--dimensional simple cubic lattice results in a shallow dip in the center of the band. 

The calculation of quantum corrections to the conductivity of a 2D lattice shows that the well--known logarithmic localization correction exists for all band fillings. Furthermore the magnitude of the correction increases as half--filling is approached. The evaluation of the obtained analytical results shows evidence for delocalized states in the center of the band of a 2D lattice with substitutional impurities. 
\end{abstract}


\section{Introduction}
The investigation of effects of weak localization in low--dimensional
electronic systems still preserves its popularity today, since
experimental studies of these systems give occasionally surprising
results like metal--insulator phase transitions in the
two--dimensional (2D) electron gas at zero magnetic
field\cite{kravchenko94a} and unusual behavior of dephasing time in
quasi one--dimensional metallic wires\cite{webb97a} at low
temperatures. Impurity effects in one--dimensional (1D) disordered
systems have been studied accurately due to the existence of
methods which give exact results for the cases of both weak and
strong disorder.\cite{lifshits88a,abrahams79a} Also perturbative
approaches give good results for three--dimensional (3D) weakly
disordered systems.

Weak localization corrections to the kinetic coefficients of 2D
disordered systems are logarithmically divergent corrections and it is
hard to sum perturbatively the leading divergent contributions. The
dimension two is the marginal dimension for the localization problem
and a small external perturbation can change the character of
localization in these systems.

According to the scaling theory of Abrahams {\it et.
  al.}\cite{abrahams79a} all states of 1D and 2D electronic gases
moving in the field of randomly distributed impurities are completely
localized irrespective of the degree of randomness, and in 3D the
Anderson metal--insulator phase transition occurs with
increasing impurity concentration. Notice that the result of the
scaling theory for 1D disordered systems is in agreement with exact
results.\cite{berezinskii73a,mott61a} The scaling theory, and also the
diagrammatic approach to the problem for 2D weakly disordered systems
has revealed logarithmic quantum corrections to the
conductivity,\cite{gorkov79a} which tend towards localization. The
logarithmic correction to the conductivity found in
[\onlinecite{abrahams79a,gorkov79a}] can be expressed as
\begin{equation}
\sigma-\sigma_0 =-\frac{e^2}{2\pi^2} \ln \frac{\tau^*}{\tau}
\label{eq:2D_log_correction}
\end{equation}
where $\sigma_0$ is the Drude conductivity, $\tau^*={\rm
  min}\{\tau_\varphi,\frac{L}{v_F},\frac{1}{\omega}\}$ with $\tau$ and
$\tau_\varphi$ being the elastic and inelastic scattering times,
respectively; $L$ is the linear size of a 2D system and $\omega$ is
the external frequency.

Notice that weak disorder gives no singular contribution to the density of
states (DOS) $\rho(\epsilon)$ of electron gas models
 when the condition $p_Fl \gg 1$ or $\epsilon_F
\tau \gg 1$ is satisfied ($p_F$ and $\epsilon_F$ are the Fermi
momentum and the Fermi energy; $l$ is the mean free path). According
to the Einstein relation $\sigma = e^2 \rho D$ the quantum correction
Eq.(\ref{eq:2D_log_correction}) to $\sigma$ is due to changes in the
diffusion coefficient\cite{gorkov79a,altshuler85a} $D$.

Even short--range and weak correlations in weakly disordered metals have been
shown\cite{altshuler85a,altshuler80b,fukuyama85a,lee85a} to result in
nontrivial quantum corrections to the DOS of low--dimensional conductors near
the Fermi level. Electron--electron interactions in disordered systems
also give rise to quantum corrections to the conductivity, which tend
to localize the electronic states. Further, the DOS of strongly doped
semiconductors in the presence of long range Coulomb interaction vanishes at the Fermi surface,\cite{pollak70a,efros75a,efros84a} which is commonly referred to as Coulomb gap.

Disordered metals in most of the papers on localization problems are
modeled as a free electron gas moving in the random field of
impurities. However, doped crystals with low concentrations of
impurities usually preserve their periodical structures and the
impurity atoms in most cases substitute the host atoms of the lattice.
Substitutional impurities appear to have strong influence on the
physical properties of low--dimensional lattices near the commensurate
values of the electron wave length $\lambda$ with the lattice constant
$a$ [\onlinecite{nakhmedov00b}].

In this paper we shall study the effects of substitutional impurities
with small concentration on the density of states and the conductivity
of a 2D square lattice. The effects of commensurability on the DOS for
the 3D simple cubic lattice will also be considered for the sake of
completeness.

Bragg reflections of the electronic wave on the Brillouin zone
boundary in the process of impurity scattering become essential as the
middle of the band is approached. This process introduces a new
relaxation time for umklapp scattering to the problem, which increases
for a deviation from half--filling due to the distortion of the
commensurability condition for the Bragg reflection.

The lattices under consideration have square structure for 2D and
simple cubic structure for 3D crystals. Our calculations show that
umklapp scattering strongly changes the energy dependence of the DOS
close to half--filling, causing it to decrease.
For the 2D square lattice the DOS of the pure system has a logarithmic
van Hove singularity at the middle of the band. This peak is shown to
be preserved for the lattice with substitutional impurities. However
its energy dependence is changed strongly, and the peak becomes
narrower.\cite{nakhmedov00b} For
the 3D lattice, the van Hove singularities lie far from the center of
the band and umklapp scattering forms a shallow dip on the Fermi
surface, the depth of which increases with the impurity concentration.

Periodicity causes the appearance of a new class of diagrams which
give a contribution to the conductivity $\sigma(\omega)$. These corrections to the
conductivity can be separated into two classes: corrections due to the
diffusion coefficient and due to the DOS. Although quantum corrections to the diffusion coefficient slightly increase $\sigma(\omega)$, the total corrections sum up to a decrease of $\sigma(\omega)$ due to DOS corrections in the vicinity of half--filling.

The paper is structured in the following form. In
Sec.\ref{sec:description} the method is described, where new diffuson
and cooperon blocks due to umklapp scattering are introduced and
calculated.  Sec.\ref{sec:dos} is devoted to the calculation of
corrections to the DOS near half--filling for 2D and 3D lattices.  In
Sec.\ref{sec:conductivity}, the calculation of the conductivity of the
2D square lattice with substitutional impurities is presented. In
Sec.\ref{sec:conclusion}, a conclusion and a discussion of the results
are given.

\section{Description of the method}
\label{sec:description}
The Hamiltonian of a $d$-dimensional simple cubic lattice with
substitutional impurities is given by
\begin{equation}
\hat{H}=\hat{H}_0+V({\bf r})
\label{eq:hamiltonian_original}
\end{equation}
where $\hat{H}_0$ is the tight--binding Hamiltonian of noninteracting
electrons in a regular lattice with lattice constant $a$ and
nearest--neighbor hopping; $V({\bf r}) = \sum_{i}U({\bf r}-{\bf R}_i)$
is the impurity potential with ${\bf R}_i$ being the positional vector
of an impurity randomly located on the $i$-th lattice site.

By introducing the second quantization operators $c_{{\bf
    p},\sigma}^\dagger$ and $c_{{\bf p},\sigma}$ for an electron with
momentum ${\bf p}$ and spin $\sigma$,
Eq.(\ref{eq:hamiltonian_original}) can be rewritten as
\begin{equation}
\hat{H}= \sum_{{\bf p},\sigma} \epsilon({\bf p}) c_{{\bf p},\sigma}^\dagger c_{{\bf p},\sigma} + \sum_{{\bf p},{\bf q},{\bf G},\sigma} \rho_{\rm imp}({\bf q}) U({\bf q}) c_{{\bf p},\sigma}^\dagger c_{{\bf p}+{\bf q}+{\bf G},\sigma}
\label{eq:hamiltonian_tight_binding}
\end{equation}

where $\rho_{\rm imp}({\bf q})=L^{-d} \sum_i \exp(i{\bf q}{\bf R}_i)$ and
$U({\bf q})$ is the Fourier transform of a single impurity potential. Throughout this
paper, we work in units of $\hbar = 1$.  The momenta ${\bf p}$ and
${\bf q}$ vary in the first Brillouin zone and ${\bf G}$ is a
reciprocal lattice vector.  $\epsilon({\bf p})$ in
Eq.(\ref{eq:hamiltonian_tight_binding}) is the
energy spectrum of an electron on a $d$-dimensional square lattice,
\begin{align}
  \epsilon({\bf p})&=t[2-\cos(p_x a)-\cos(p_y a)] &\mbox{for} \quad
  d=2
\label{eq:energy_2D}\\
\epsilon({\bf p})&=t[3-\cos(p_x a)-\cos(p_y a)-\cos(p_z a)]
&\mbox{for} \quad d=3\label{eq:energy_3D}
\end{align}
where $p_{x,y,z}=\frac{2\pi}{a N_{x,y,z}} n_{x,y,z}$ with
$\frac{-N_{x,y,z}}{2}<n_{x,y,z}\leq \frac{N_{x,y,z}}{2}$ and $t$ is
the tunneling integral for nearest--neighbor sites.

\begin{figure}
  \centerline{\epsfbox{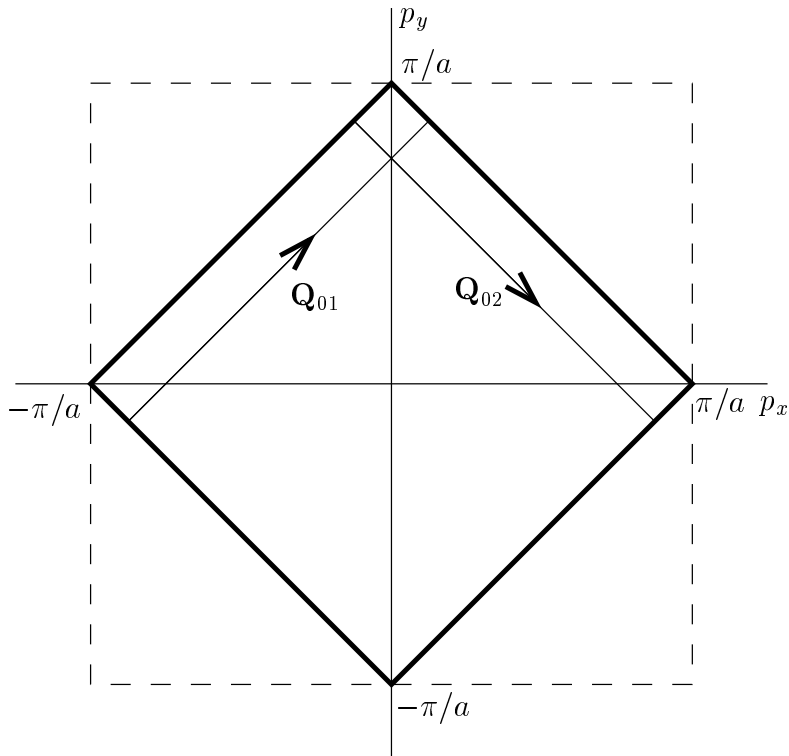}}
\caption{Fermi surface of a square lattice with nearest--neighbor hopping at half--filling. Complete nesting occurs with the nesting vectors ${\bf Q}_0=\pm{\bf Q}_{01};\pm{\bf Q}_{02}$, where ${\bf Q}_{01}=\{\frac{\pi}{a},\frac{\pi}{a}\}$ and ${\bf Q}_{02}=\{-\frac{\pi}{a},\frac{\pi}{a}\}$ are shown in the figure.}
\label{fig:fs_square}
\end{figure}

The electronic bandwidth is $W=4t$ for $d=2$ and $W=6t$ for 3D
systems.  As half--filling is approached, which corresponds to
$\epsilon_F=2t$ for 2D and $\epsilon_F=3t$ for 3D system, the Fermi
surface becomes nested, i.e. it exhibits surface
elements which can be mapped onto each other through one nesting vector.
In our case, the nesting condition is satisfied for all parts of the surface, and
for 2D square lattices, the Fermi surface at half--filling is flat
(see Fig.\ref{fig:fs_square}).

The second part of the Hamiltonian $\hat{H}$ in
Eq.(\ref{eq:hamiltonian_tight_binding}), which describes an electron
scattering on randomly distributed impurities, contains both normal
(for ${\bf G}=0$) and umklapp (for ${\bf G}\neq 0$) processes. The
impurity concentration is assumed to be small, so that the Born
approximation is appropriate\cite{abrikosov63a} to estimate the
scattering process on the impurities. The impurity potential in this
case is chosen to be a $\delta$-correlated Gaussian potential with
zero average value.

Conventional diagrammatic techniques \cite{abrikosov63a} are applied
to calculate the effects of umklapp scattering on the DOS and the
conductivity. The bare Green's functions $G_{R,A}^0(\epsilon,{\bf p})$
at zero temperature are given as
\begin{equation}
G_{R,A}^0(\epsilon,{\bf p})=\frac{1}{\epsilon-(\epsilon({\bf p})-\epsilon_F)+\frac{i}{2\tau} {\rm sign} \epsilon}
\label{eq:bareGF}
\end{equation}

where $\epsilon>0$ and $\epsilon<0$ correspond to retarded $G_R^0$ and
advanced $G_A^0$ Green's functions, respectively. The energy spectrum
$\epsilon({\bf p})$ in Eq.(\ref{eq:bareGF}) is expressed by
Eqs.(\ref{eq:energy_2D}) and (\ref{eq:energy_3D}) for 2D and 3D cases,
respectively. For momenta lying close to the Fermi surface (in the
first Brillouin zone) the energy spectrum can be linearized around the
Fermi surface and $\epsilon({\bf p})-\epsilon_F\approx {\bf v}_F({\bf
  p}-{\bf p}_F)=v_F(|{\bf p}|-p_F)\cos \alpha$, where ${\bf v}({\bf
  p}) = \frac{\partial \epsilon}{\partial {\bf p}} = a t\{\sin p_x a,
\sin p_y a\}$ is the group velocity of the electron wave packet. In
contrast to the electron gas model the linearized energy spectrum
contains the additional parameter $\cos \alpha$ due to
non--collinearity of the momentum and velocity vectors.  For small
band filling, the Fermi surface is very similar to a sphere, so this
factor can be set equal to unity. Approaching half--filling, this
parameter becomes weakly varied, e.g. $\frac{1}{\sqrt{2}}\leq \cos
\alpha \leq 1$ for a 2D system at half--filling, where the range of
variation of $\cos \alpha$ is maximal. As an approximation, we can
always set $\cos \alpha = 1$. In the vicinity of commensurate points,
especially at half--filling, electron scattering on impurities with
large momentum transfer involves umklapp processes. The procedure of
expansion of the energy spectrum around the Fermi surface can in this
case be performed after the separation of the large momentum transfer
${\bf Q}_0$.

\begin{figure}
  \centerline{\epsfbox[235 171 625 669]{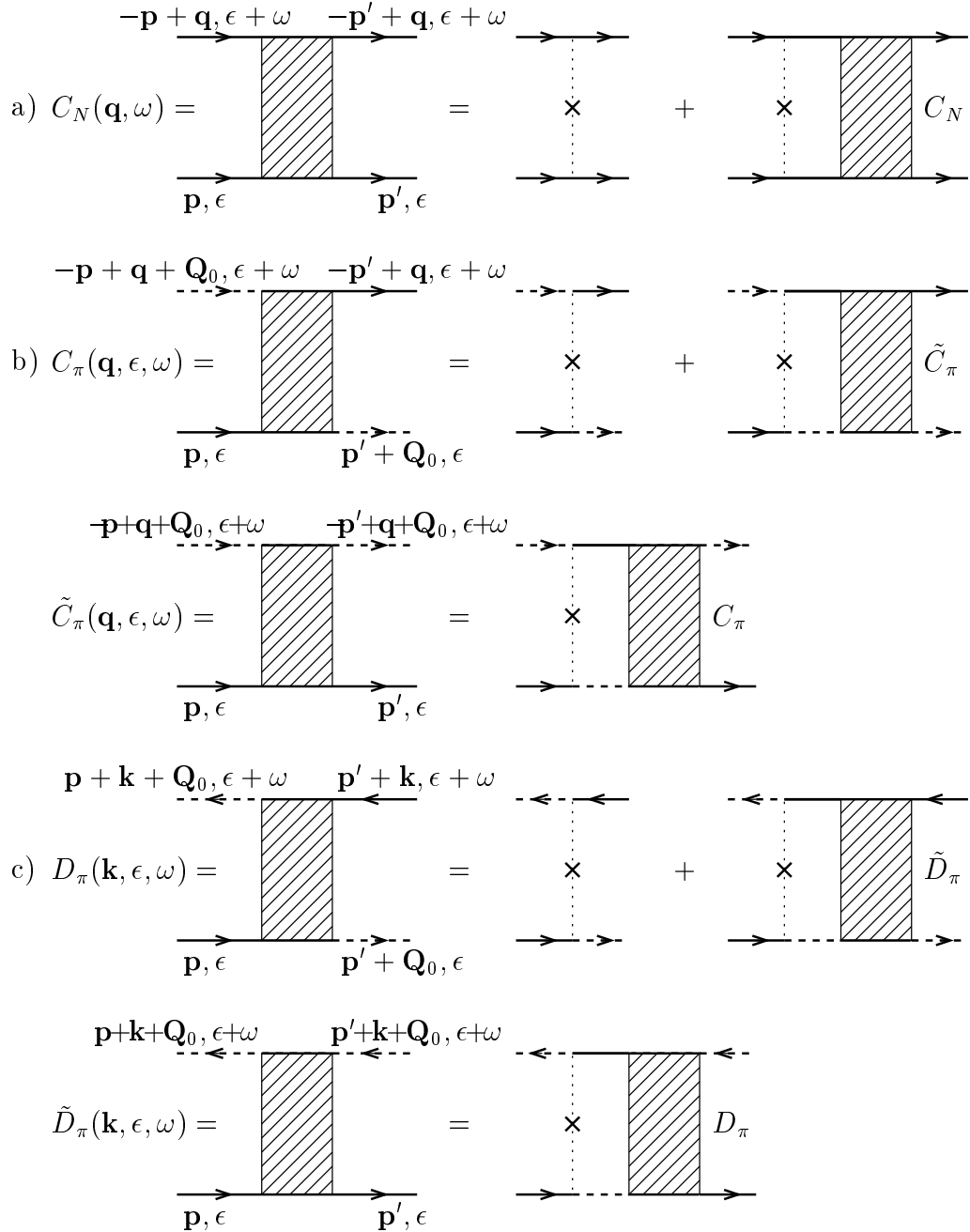}}
\caption{The ladder series for a) the normal cooperon $C_N$, b) the $\pi$--cooperon $C_\pi$ and c) the $\pi$--diffuson $D_\pi$. In b) and c), $\tilde{C}_\pi$ and $\tilde{D}_\pi$ are expressed by $\tilde{C}_\pi = J_\pi C_\pi$ and $\tilde{D}_\pi = J_\pi  D_\pi$, where $J_\pi $ is given in Eq.(\ref{eq:Ju}).}
\label{fig:ladder_diagrams}
\end{figure}

In weak localization theory the maximally crossed diagrams are
responsible for the low temperature quantum corrections to the kinetic
properties of low--dimensional disordered systems, modeled as an
electron gas moving in the field of randomly distributed impurities
\cite{gorkov79a}. These diagrams can be redrawn as ladder diagrams in
a particle--particle channel (Fig.\ref{fig:ladder_diagrams}a).  The
cooperon block has a pole for small total momentum of particles,
$|{\bf q}|l\ll 1$, and for small energy difference $|\omega|\tau \ll
1$. The cooperon block turns out to have a diffusion pole also for
large momentum transfer when the total momenta of the particles are
$({\bf Q}_0+{\bf q})$ with $|{\bf q}|l\ll 1$ and when the total energy is
small.  In this case each act of scattering on an impurities 
involves a large momentum transfer close to ${\bf Q}_0$.
This process of course takes place mainly around half--filling when
${\bf Q}_0=2{\bf p}_F$, and it consists of simultaneous Bragg
reflection of the electron in the process of scattering on the
impurity, with relaxation time $\tau_\pi$. This scattering process
will be further referred to as $\pi$--scattering.

\begin{figure}
  \centerline{\epsfbox[280 424 640 669]{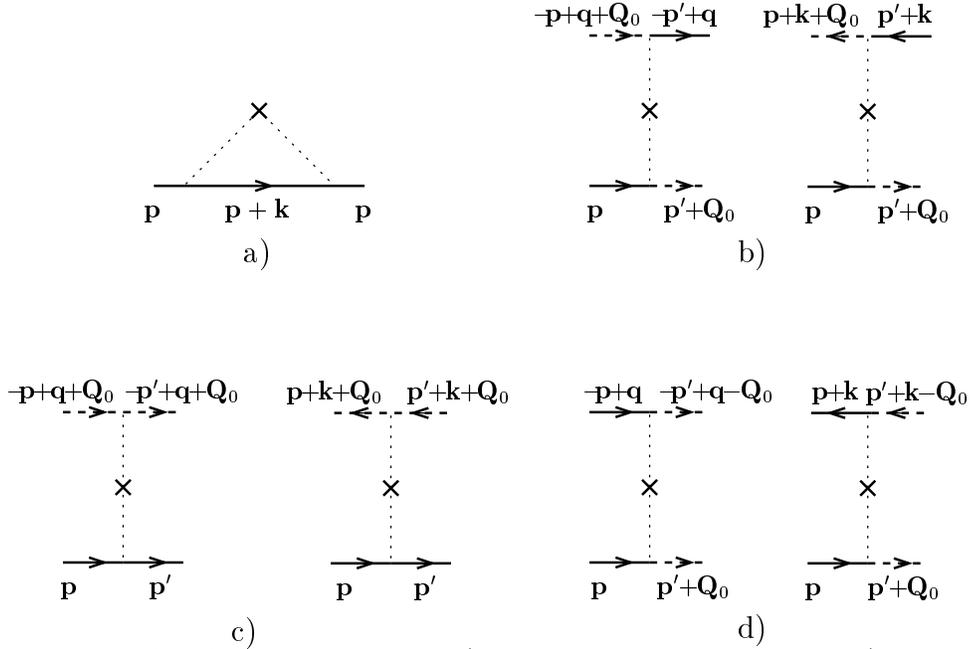}}
\caption{Impurity vertices which give a contribution to a) the normal relaxation time $\tau_N$, b) the relaxation time for $\pi$--scattering $\tau_\pi$. The vertices c) and d) do not give a contribution to $\pi$--scattering.}
\label{fig:vertices}
\end{figure}

Two different scattering processes and corresponding relaxation times
can be separated: Normal scattering with $\tau_N$ (see
Fig.\ref{fig:vertices}a), when the scattering on an impurity maintains
the electron's momentum inside the first Brillouin zone
after scattering and $\pi$--scattering, with a relaxation time
$\tau_\pi$ (Fig.\ref{fig:vertices}b). Here, momentum conservation is
violated and this scattering process corresponds to an umklapp
process. The total relaxation time $\tau$ can then be expressed as
$\frac{1}{\tau}=\frac{1}{\tau_N}+\frac{1}{\tau_\pi}$.  Notice here
that the expression for the normal relaxation time $\tau_N$ according
to Fig.\ref{fig:vertices}a for the 2D case can be given as
\begin{equation}
\frac{1}{\tau_N}
=\frac{C_{\rm imp}}{(2\pi)^2} 
\int_S \frac{{\rm d} {\bf S}}{|{\bf v_k}|}|U({\bf p}_F,{\bf S})|^2
\end{equation}
We assume $\tau_N$ to be constant and independent of the band filling.
This means that the singularities of the integrand due to vanishing
velocity at the saddle points are compensated by appropriate zeros of
the impurity potential.

To illustrate diagrammatically the impurity scattering including
umklapp processes we represent the Green's function of an electron
with large momentum by a dashed line.  The ladder series for the
cooperon block with large momentum transfer, $C_\pi({\bf
  q},\epsilon)$, is diagrammatically shown in
Fig.\ref{fig:ladder_diagrams}b. Notice here that the vertices c) and
d) in Fig.\ref{fig:vertices} are irrelevant for the cooperon blocks
$C_\pi({\bf q},\epsilon)$ and $C_N({\bf q},\epsilon)$ (Fig.
\ref{fig:ladder_diagrams}a,b). Indeed, the dashing of the line alone
has no physical meaning, it has only meaning for an electron
scattering with large momentum transfer. In ladders, the vertices c) and d) of Fig.\ref{fig:vertices} can always be transformed into normal vertices or those of Fig.\ref{fig:vertices}b by proper redefinition of the internal momenta.

In the process of propagation in the particle--particle channel the
momenta of the particles for the normal cooperon block $C_N({\bf
  q},\omega)$ lie on opposite sites of the Fermi surface during the
diffusion process. In contrast to this, the momenta for the two
particle propagator $C_\pi ({\bf q},\epsilon)$ with umklapp process
lie on the same section of the Fermi surface and each act of impurity
scattering coherently relocates the particles to the opposite section
of the Fermi surface due to Bragg reflection. We will refer to this
two--particle propagator $C_\pi ({\bf q},\epsilon)$ as $\pi$-cooperon.

The same situation occurs for the $\pi$--diffusion block $D_\pi({\bf
  k},\epsilon,\omega)$ (Fig.\ref{fig:ladder_diagrams}) with a large
difference  (${\bf Q}_0+{\bf k}$) of electron and hole momenta, with $|{\bf k}|l \ll 1$. The
comparison of the normal diffuson block $D_N({\bf k},\epsilon,\omega)$
with the $\pi$--diffuson $D_\pi({\bf k},\epsilon,\omega)$ reveals
again strong differences between these two processes. In the process
of propagation in the $N$--diffusion channel , electron and hole lie
close to each other on the Fermi surface, with momenta ${\bf p}$ and
${\bf p}+{\bf k}$, and $|{\bf k}|l\ll1$. They diffuse on the Fermi
surface and keep the proximity of the momenta after each act of
scattering. In contrast to this, the momenta of electron and hole for
the $\pi$--diffuson block $D_\pi$ are located on opposite sides of the
Fermi surface. Each scattering on the impurities interchanges the
positions of electron and hole on the Fermi surface. Now let us
sum the series shown in Fig.\ref{fig:ladder_diagrams} for $C_N$,
$C_\pi$, and $D_\pi$.

Summing up the ladder series in Fig.\ref{fig:ladder_diagrams}a, the
following expression for $C_N({\bf q},\omega)$ is obtained:
\begin{equation}
C_N({\bf q},\omega)=C_{\rm imp} |U(0)|^2\Bigl\{
\frac{\sqrt{(1-i\tau|\omega|)^2+(ql)^2}}{\sqrt{(1-i\tau|\omega|)^2+(ql)^2}-\frac{\tau}{\tau_N}}\Theta(-\epsilon(\epsilon+\omega)) + \Theta(\epsilon(\epsilon+\omega))\Bigr\}
\label{eq:Cn}
\end{equation}
Far from half--filled electronic band, $\tau_\pi \gg \tau_N$ and $\tau
\approx \tau_N$. Therefore, $C_N({\bf q},\omega)$ has a diffusion pole
for $|\omega|\tau \ll 1$ and $|{\bf q}|l \ll 1$. As half--filling is
approached, umklapp scattering is intensified and
$\tau_\pi \ll \tau_N$ and $\tau \approx \tau_\pi$. As a result, the
diffusion pole of $C_N({\bf q},\epsilon)$ disappears at half--filling.
In this regime, the Fermi surface of the square or simple cubic
crystal becomes nested (see Fig.\ref{fig:fs_square} for the 2D case)
with a nesting vector ${\bf Q}_0=\{\pm \frac{\pi}{a},\pm
\frac{\pi}{a}\}$ for the 2D lattice and ${\bf Q}_0=\{\pm
\frac{\pi}{a},\pm \frac{\pi}{a},\pm \frac{\pi}{a}\}$ for the 3D
lattice. In this case, the following particle--hole symmetry
of the electron dispersion with respect to the vector ${\bf Q}_0$
holds for the half--filled band:
\begin{equation}
\epsilon({\bf p}+{\bf Q}_0)-\epsilon_F = - [\epsilon({\bf p})-\epsilon_F]
\label{eq:nesting}
\end{equation}
The main contribution to the low temperature properties now gives the
$\pi$--cooperon $C_\pi ({\bf q},\epsilon,\omega)$ in
Fig.\ref{fig:ladder_diagrams}b. One gets the following expression for
$C_\pi ({\bf q},\epsilon,\omega)$ by summing the ladder series in
Fig.\ref{fig:ladder_diagrams}:
\begin{equation}
C_\pi ({\bf q},\epsilon,\omega) = 
C_{\rm imp}|U|^2 
\frac{J_\pi ^2}{1-J_\pi ^2} 
\label{eq:Cp}
\end{equation}

where $J_\pi $ is the expression for an elementary ``bubble'' in the
ladder series and
\begin{equation}
J_\pi =C_{\rm imp} \int \frac{{\rm d}^d p}{(2 \pi)^d} |U|^2 G({\bf p},\epsilon),G(-{\bf p}+{\bf q}+{\bf Q}_0,\epsilon+\omega)
\label{eq:Ju}
\end{equation}
To calculate $J_\pi $, the large momentum ${\bf Q}_0$ is
removed using the electron--hole symmetry relation
Eq.(\ref{eq:nesting}), and after this the energy spectrum is
linearized around the Fermi surface. As a result we get the following
expression for $C_\pi ({\bf q},\epsilon,\omega)$:
\begin{equation}
C_\pi ({\bf q},\epsilon,\omega) = C_{\rm imp}|U(2 p_F)|^2 
\Bigl\{ \frac{(\tau/\tau_\pi)^2}{(1-i\tau|2\epsilon + \omega|)^2 + (ql)^2-(\tau/\tau_\pi)^2} \Theta(\epsilon(\epsilon+\omega)) + \Theta(-\epsilon(\epsilon+\omega))\Bigr\}
\label{eq:Cp_sol}
\end{equation}

So the $\pi$--cooperon has a diffusion pole for total momenta close to
${\bf Q}_0$ ($|{\bf k}|l \ll 1$) and small total energy
$|2\epsilon+\omega|\tau \ll 1$ of the particles.  The diffusion block
$D_\pi ({\bf k},\epsilon,\omega)$, which 
differs from  $C_\pi ({\bf q},\epsilon,\omega)$ through time reversal of one electron line, has a pole for large (${\bf Q}_0 + {\bf k}$) momenta difference with $|{\bf k}|l \ll 1$ and small total energy of electron and
hole (see Fig.\ref{fig:ladder_diagrams}c).  The calculation of $D_\pi
({\bf k},\epsilon,\omega)$ is similar to that for $C_\pi ({\bf
  q},\epsilon,\omega)$ and we obtain
\begin{equation}
D_\pi ({\bf k},\epsilon,\omega) = C_{\rm imp}|U(2 p_F)|^2 
\Bigl\{ \frac{(\tau/\tau_\pi)^2}{(1-i\tau|2\epsilon + \omega|)^2 + (kl)^2-(\tau/\tau_\pi)^2} \Theta(\epsilon(\epsilon+\omega)) + \Theta(-\epsilon(\epsilon+\omega))\Bigr\}
\label{eq:Dp_sol}
\end{equation}
After this, we can calculate the corrections to the DOS and the
conductivity due to umklapp processes.

\section{Density of States at Half--filling}
\label{sec:dos}
The one--particle DOS of the regular $d$--dimensional lattice is
expressed as
\begin{equation}
\rho_0^{(d)}=\frac{2}{(2\pi)^d} 
\int_S \frac{{\rm d} {\bf S}}{|{\bf \nabla} \epsilon({\bf k})|}
\end{equation}
where ${\rm d}{\bf S}$ is an element of an isoenergetic surface in
$d$--dimensional space. $\rho_0^{(d)}$ has a van Hove singularity at
the points where the group velocity of the electron wave packet ${\bf
  v}_k = {\bf \nabla}\epsilon({\bf k})$ vanishes.\cite{ashcroft}

The DOS of a clean 1D lattice with constant spacing $a$ is easily
calculated to be $\rho_0^{(1)} = \frac{1}{\pi a \sqrt{\epsilon(2t -
    \epsilon)}}$. This expression shows that $\rho_0^{(1)}$ is a
regular function of the energy near half--filling when $\epsilon \to
\epsilon_F = t$ and it has a power--like singularity when $\epsilon$
approaches the band edges $\epsilon \to 0,2t$. The bare DOS of a 2D
crystal with an energy spectrum given by Eq.(\ref{eq:energy_2D}) has a
logarithmic van Hove singularity in the middle of the band:
\begin{equation}
\rho_0^{(2)} = \frac{1}{\pi^2 a^2 t} K\bigl(\sqrt{1-(\frac{\epsilon}{2t})^2}\bigr) =
\begin{cases}
  \frac{2}{\pi a^2 \sqrt{4 t^2-\epsilon^2}}\ln \frac{4 t^2-\epsilon^2}{\epsilon^2} & \mbox{for $|\epsilon| \ll 2t$}\\
  \frac{1}{\pi a^2|\epsilon|} &\mbox{for $|\epsilon| \approx 2t$}
\end{cases}
\label{eq:rho_2D}
\end{equation}
Here and below, 
the electron energy $\epsilon$ is measured from the middle of the
band, i.e. $\epsilon=0$  corresponds to half--filling.  The van
Hove singularity of simple 3D crystals with nearest--neighbor hopping
exhibits cusps near $|\epsilon|=t$:
\begin{equation}
\rho_0^{(3)}={\rm const}- \frac{2}{\pi^2 t^{3/2} a^3}\sqrt{\epsilon-t}
\label{eq:rho_3D}
\end{equation}

In this section we will study effects of substitutional impurities on
the DOS of 2D and 3D crystals. Randomly distributed impurities in
$d$--dimensional electron gases have no effect on the DOS. 

The DOS of 2D and 3D simple (cubic) crystals with substitutional
impurities turns out to have a quantum correction near the middle of the band
even for noninteracting electrons. Effects of commensurability on the
DOS and on the kinetics of 1D disordered crystals near the middle of
the band have been calculated by many
authors.\cite{dyson53a,weissman75a,gorkov76a,ovchinnikov77a,gogolin77a,gredeskul78a,hirsch76a,eggarter78a,inui94a}
Dyson first pointed out\cite{dyson53a} that the DOS of phonons of a 1D
disordered chain has a singularity as $\rho^{(1)}\propto
-|\epsilon|^{-1} \ln^{-3}|\epsilon|$ near the middle of the band.
Notice that the singular increasing of the DOS in a 1D system at
half--filling is not connected with the van Hove singularity, since
the latter is located at the band edge for the 1D system. Instead it is
mediated by the interference of impurity scattering and Bragg
reflections at half--filling. Later an analogous singularity has been
found in the electronic DOS of many 1D
models.\cite{weissman75a,gorkov76a,ovchinnikov77a,gogolin77a,gredeskul78a,hirsch76a,eggarter78a,inui94a}
Since the 1D scattering problem is characterized by forward and
backward scattering processes, 1D models with off--diagonal disorder
display a singularity at the center of the band only if the forward
scattering amplitude turns to zero. Small forward scattering
with Bragg reflection seems to enhance localization and as a result the DOS
divergence in the band center is blurred.

Effects of substitutional impurities on the DOS of 2D (square,
honeycomb and triangular) and 3D (simple cubic) lattices have been
studied computationally\cite{hu84a,schreiber98a,kramer93} for cases
with diagonal and off--diagonal disorder.
The existence of a van Hove singularity in the DOS of a two sublattice
model for $d\geq2$ has been shown in [$\onlinecite{oppermann79a}$] on
the basis of an $\frac{1}{n}$ expansion in disordered systems with
$n$ orbitals per site for energies approaching the band center.
In all cases diagonal disorder has been shown to suppress the van Hove
singularity, whereas it is preserved for the 2D simple square lattice
with off-diagonal disorder.

\begin{figure}
  \centerline{\epsfbox[239 423 591 670]{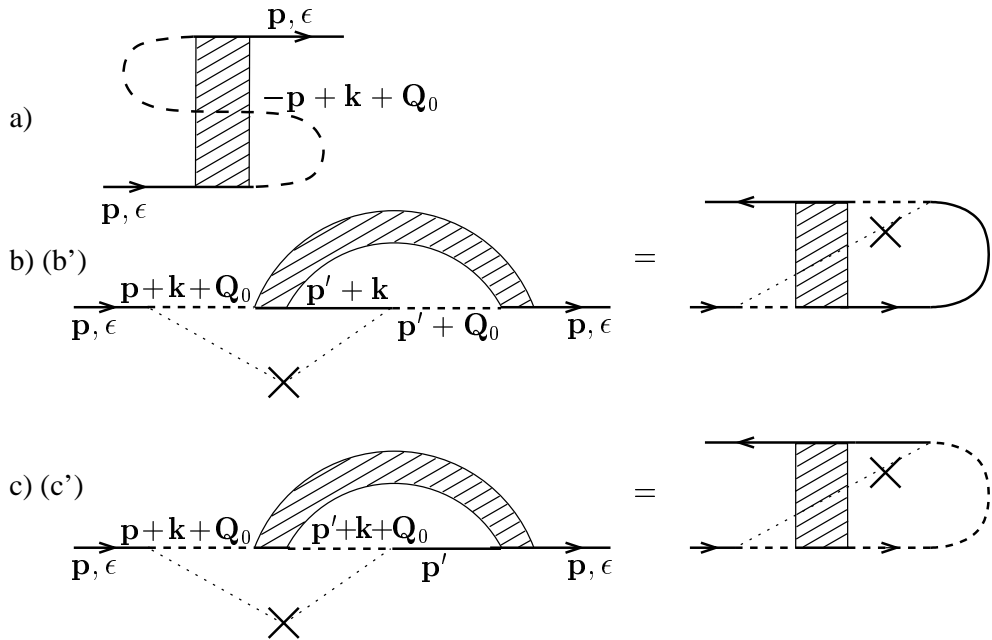}}
\caption{Self energy parts which give the leading contributions to the Green's function due to umklapp scattering in a) cooperon and b,c) diffuson channels. Diagrams symmetrically conjugate to b) and c) with respect to the impurity line, also give a contribution to the self--energy.}
\label{fig:self_energy_diagrams}
\end{figure}

 We start from the following expression
for the DOS
\begin{equation}
\rho^{(d)}(\epsilon)=-\frac{2}{\pi} {\rm Im} \int \frac{{\rm d}^d p}{(2\pi)^d} G_{R}({\bf p},\epsilon)
\label{eq:rho_from_gf}
\end{equation}
which expresses $\rho(\epsilon)$ by means of the retarded Green's
function $G_{R}({\bf p},\epsilon)$. The new class of diagrams which
give the dominating contribution to the self--energy of $G_R({\bf
  p},\epsilon)$ is drawn in Fig.\ref{fig:self_energy_diagrams}.

 These diagrams represent the contributions to the self--energy in
first order in cooperon and diffuson blocks ($C_\pi({\bf q},\epsilon)$
and $D_\pi({\bf k},\epsilon)$). Since $C_\pi({\bf q},\epsilon)$ and
$D_\pi({\bf k},\epsilon)$ do not carry an external frequency ($\omega
= 0$), their expressions are obtained from Eqs.(\ref{eq:Cp_sol}) and
(\ref{eq:Dp_sol}) with $\omega=0$:

\begin{align}
  C_\pi ({\bf q},\epsilon) &= C_{\rm imp}|U(2 p_F)|^2
  \bigl(\frac{\tau}{\tau_\pi}\bigr)^2
  \frac{1}{(1-2 i\tau|\epsilon|)^2 + (q l)^2-(\tau/\tau_\pi)^2} \label{eq:Cp0}\\
  D_\pi ({\bf k},\epsilon) &= C_{\rm imp}|U(2 p_F)|^2
  \bigl(\frac{\tau}{\tau_\pi}\bigr)^2 \frac{1}{(1-2 i\tau|\epsilon|)^2
    + (k l)^2-(\tau/\tau_\pi)^2}
  \label{eq:Dp0}
\end{align}
As $\tau_\pi \to \tau$, both blocks have a diffusion pole for
$|\epsilon|\tau \ll 1$ and $ql,kl \ll 1$.  The cooperon and the
diffuson blocks give rise to logarithmic corrections to the physical
parameters for $d=2$ and $\sqrt{|\epsilon|\tau}$ corrections for
$d=3$.  Therefore, diagrams of higher order in the cooperon and
diffuson blocks for the self--energy parts are not necessary for 3D
systems for $|\epsilon|\tau \ll 1$. Nevertheless the logarithmic
divergency existing in the first order contribution to the
self--energy for 2D systems requires to examine higher order
logarithmic corrections. The structure of the diagrams which contain
all possible combinations of diffuson and cooperon blocks becomes
complicated with increasing number of inserted blocks.

However, as it has been shown by various theoretical methods, higher
logarithmic corrections to the conductance cancel each
other\cite{gorkov79a,altshuler86b,oppermann79a} and the leading
correction to the conductance is only the first order logarithmic
term.

\begin{figure}
  \centerline{\epsfbox[175 232 386 668]{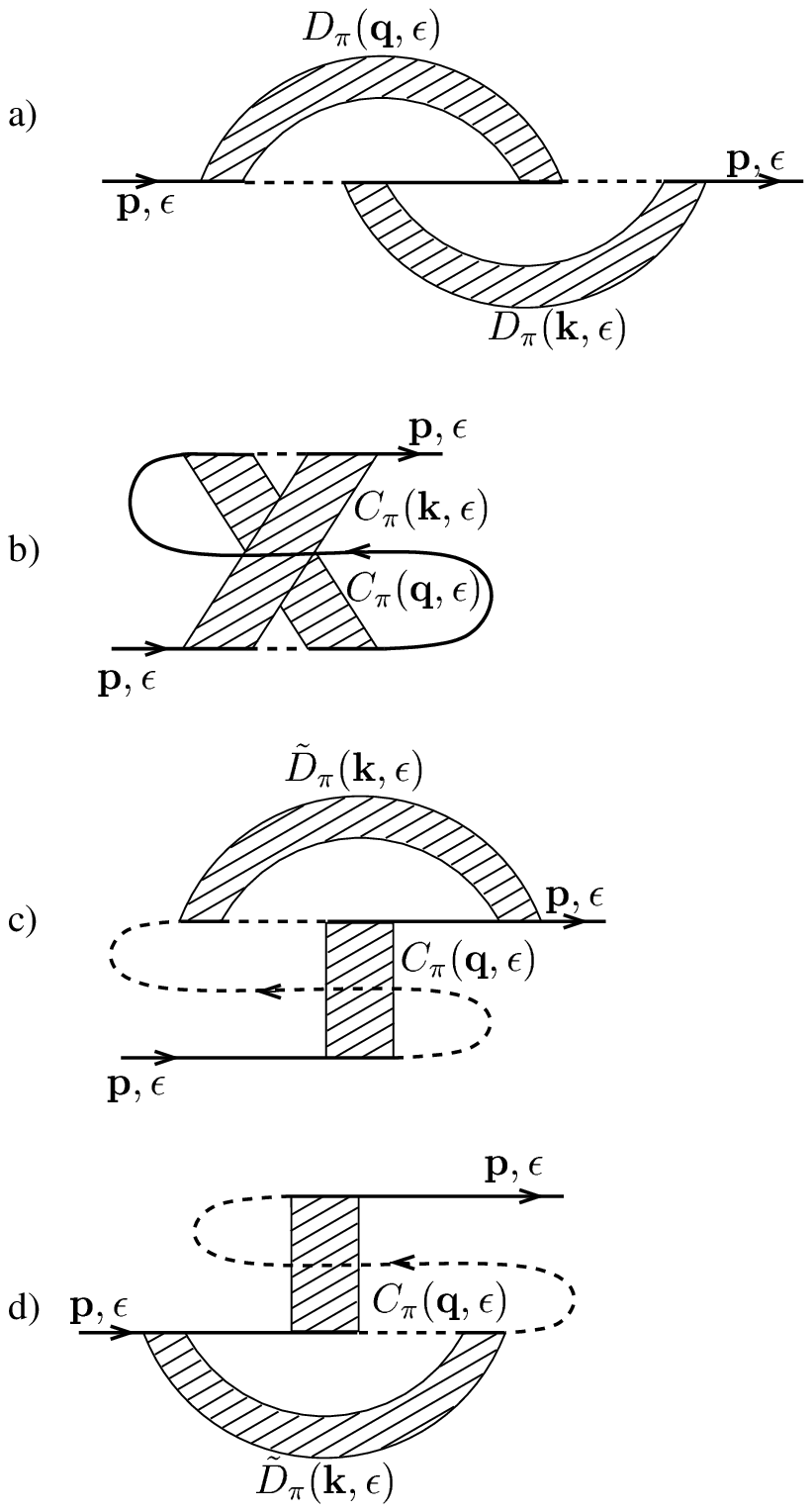}}
\caption{Contributions of second order in cooperon and diffuson blocks to the self--energy. The calculations shows a complete cancellation between these diagrams. The second--order self--energy diagrams obtained by mutual insertion of the diagrams of Fig.\ref{fig:self_energy_diagrams}, with a total number of 40, also completely cancel each other at half--filling.}
\label{fig:self_energy_second_order}
\end{figure}

High order corrections to the self--energy seem to cancel each other
also in our case. The second order diagrams which give
logarithm--squared contributions to the self--energy are drawn in
Fig.\ref{fig:self_energy_second_order}. Straightforward calculations
show that the sum of these diagrams gives zero.

The retarded Green's function $G_R(\epsilon,{\bf p})$ is expressed by
the sum of the first--order self--energy parts $\Sigma(\epsilon,{\bf
  p})$ in Fig.\ref{fig:self_energy_diagrams} according to the Dyson
equation as
\begin{equation}
G_K(\epsilon,{\bf p})=\frac{1}{(G_R^0(\epsilon,{\bf p}))^{-1}-\Sigma(\epsilon,{\bf p})}
= \sum_{n=0}^\infty \bigl(G_R^0(\epsilon,{\bf p})\bigr)^{n+1}\bigl(\Sigma(\epsilon,{\bf p})\bigr)^n
\label{eq:fullGF}
\end{equation}
By summing the diagrams in Fig.\ref{fig:self_energy_diagrams} one gets
the expression for $\Sigma(\epsilon,{\bf p})$:
\begin{equation}
\Sigma(\epsilon,{\bf p})=\int \frac{{\rm d}^d k}{(2 \pi)^d} \bigl\{ C_\pi(\epsilon,{\bf k})-\frac{2 \tau}{\tau_\pi}\bigl(1-\frac{\tau}{\tau_\pi}\bigr) D_\pi(\epsilon,{\bf k})\bigr\} G_R^0(-{\bf p}+{\bf k}+{\bf Q}_0,\epsilon)
\label{eq:sigma}
\end{equation}
By substituting Eqs.(\ref{eq:fullGF}) and (\ref{eq:sigma}) into
Eq.(\ref{eq:rho_from_gf}) one obtains the
following expression for $\rho(\epsilon)$:
\begin{equation}
\rho(\epsilon)=\rho_0^{(d)}-\frac{2}{\pi} {\rm Im} \sum_{n=1}^\infty A_n \alpha_d^n(\epsilon)
\label{eq:rho_as_sum}
\end{equation}
with
\begin{equation}
\alpha_d(\epsilon)=4\tau^2\int \frac{{\rm d}^dk}{(2\pi)^d} \bigl\{ 
C_\pi(\epsilon,{\bf k})-\frac{2\tau}{\tau_\pi}\bigl(1-\frac{\tau}{\tau_\pi}\bigr) D_\pi(\epsilon,{\bf k})\bigr\}
\label{eq:alpha}
\end{equation}
and
\begin{equation}
A_n=\frac{1}{(2\tau)^{2n}} \int \frac{{\rm d}^dk}{(2\pi)^d} 
G_0^{n+1}(\epsilon,{\bf p})G_0^n(-{\bf p}+{\bf p}_0,\epsilon)=(-1)^{n+1}2\pi i \rho_0^{(d)}\frac{n(2n-1)!}{2^{2n}(n!)^2}
\label{eq:An} 
\end{equation}
Here, $\rho^{(d)}_0$ is the DOS for a $d$--dimensional clean crystal
and is given by Eqs.(\ref{eq:rho_2D}) and (\ref{eq:rho_3D}) for $d=2$
and $d=3$, respectively.  The sum over $n$ in Eq.(\ref{eq:rho_as_sum})
is performed easily using Eq.(\ref{eq:An}):
\begin{equation}
\rho(\epsilon)=\rho_0^{(d)}\bigl\{1-{\rm Re} \frac{\alpha_d(\epsilon)}{\sqrt{1+\alpha_d(\epsilon)}(1+\sqrt{1+\alpha_d(\epsilon)})}\bigr\}=\rho_0^{(d)}{\rm Re}\frac{1}{\sqrt{1+\alpha_d(\epsilon)}}
\label{eq:rho_from_alpha}
\end{equation}
where the energy dependence of $\alpha_d(\epsilon)$ is obtained from
Eqs.(\ref{eq:alpha}), (\ref{eq:Cp0}) and (\ref{eq:Dp0}):

\begin{equation}
\alpha_d(\epsilon)=
\begin{cases}
  \frac{1}{2\pi \epsilon_F \tau_\pi}\bigl(1-\frac{2 \tau}{\tau\pi}+\frac{2 \tau^2}{\tau_\pi^2}\bigr) \ln \frac{2-\frac{\tau^2}{\tau_\pi^2}}{1-\frac{\tau^2}{\tau_\pi^2}-4 i |\epsilon| \tau} & \mbox{for $d=2$}\\
\begin{split}
  \frac{3
    \bigl(1-\frac{2\tau}{\tau\pi}+\frac{2\tau^2}{\tau_\pi^2}\bigr)}{4 \pi
    \tau \tau_\pi \epsilon_F^2}\Bigl\{1-\frac{\pi
    \sqrt{3}}{4}\bigl[\sqrt{\sqrt{(1-\frac{\tau^2}{\tau_\pi^2})^2+(4\tau
      |\epsilon|)^2}-1+\frac{\tau^2}{\tau_\pi^2}}\\
  -i \sqrt{\sqrt{(1-\frac{\tau^2}{\tau_\pi^2})^2+(4\tau
      |\epsilon|)^2}+1-\frac{\tau^2}{\tau_\pi^2}}\bigr]\Bigr\}\end{split}
&\mbox{for $d=3$}
\end{cases}
\label{eq:alpha_sol}
\end{equation}
It can be seen from Eqs.(\ref{eq:rho_from_alpha}) and
(\ref{eq:alpha_sol}) that away from half--filling, the effects of
Umklapp scattering are weakened and $\tau_\pi \gg \tau_N \approx
\tau$, as a result of which the quantum corrections to the DOS
disappear. In the vicinity of half--filling, $\tau\approx \tau_\pi <
\tau_N$ and impurity effects become essential.

The DOS for a 2D system with half--filled energy band can be expressed
as
\begin{equation}
\rho^{(2)}(\epsilon)=\rho_0^{(2)}(\epsilon)\frac{1}{\sqrt{1+\frac{1}{2\pi \epsilon_F\tau_\pi}\ln\frac{1}{4 |\epsilon| \tau_\pi}}}
\label{eq:rho_2D_sol}
\end{equation}
where $\rho_0^{(2)}(\epsilon)$, given by Eq.(\ref{eq:rho_2D}),
contains a logarithmic singularity in the middle of the band.
Eq.(\ref{eq:rho_2D_sol}) shows that the van Hove singularity in the
DOS of a pure 2D square crystal is preserved in the presence of
substitutional impurities. However, the central peak becomes narrower
due to impurity scattering than that of the  van Hove peak in clean systems. In
the vicinity of the band center the energy dependence of
$\rho^{(2)}(\epsilon)$ is changed from logarithmic dependence to the
square root of the logarithm,\cite{nakhmedov00b}
\begin{equation}
\rho^{(2)}(\epsilon)=\frac{2}{(\pi a)^2 \epsilon_F}(2\pi \epsilon_F \tau_\pi)^{1/2}\ln^{1/2}(\frac{1}{4\tau_\pi |\epsilon|}) \quad \mbox{as} \quad |\epsilon| \to 0
\label{eq:rho_2D_singularity}
\end{equation}

\begin{figure}
  \centerline{\epsfbox[183 562 516 669]{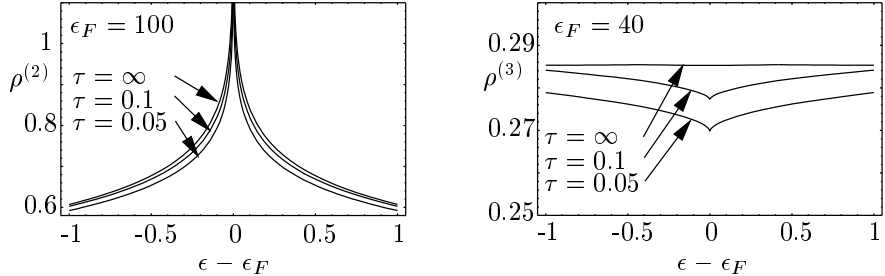}}
\caption{Impurity effect on $\rho^{(2)}$ and $\rho^{(3)}$ for selected values of the impurity potential. The pure DOS $\rho_0^{(2)}$ and $\rho_0^{(3)}$, corresponding to $\tau=\infty$, are also given for comparison.}
\label{fig:rho_2D}
\end{figure}

In Fig.\ref{fig:rho_2D}a the dependence of $\rho^{(2)}(\epsilon)$ on
the impurity potential strength is drawn, which displays a narrowing
of the central peak with increasing disorder strength or decreasing
$\tau_\pi$. The DOS of a square lattice with off--diagonal disorder
computed in [\onlinecite{hu84a,schreiber98a}] shows the same effect.
Notice that the impurity potential studied in our problem corresponds
to off--diagonal disorder, since the part of the Hamiltonian
Eq.(\ref{eq:hamiltonian_tight_binding}) corresponding to the impurity
potential can be decoupled into diagonal ($U({\bf 0})$) and
off--diagonal ($U({\bf q})$ with ${\bf q}\neq0$) parts. The diagonal
part is chosen to be equal to zero for the white--noise
potential\cite{abrikosov63a} in the averaging process.

Unlike 2D systems, substitutional
impurities have small effect on the DOS of 3D simple cubic crystals.
In the close vicinity of half--filling, Eq.(\ref{eq:alpha}) for
$\alpha_3(\epsilon)$ is simplified:
\begin{equation}
\alpha_3(\epsilon)=\frac{3}{4 \pi \tau_\pi^2 \epsilon_F^2} \bigl\{1-\frac{\pi\sqrt{3}}{2}\sqrt{|\epsilon|\tau}(1-i)\bigr\} \equiv \alpha_R + i \alpha_I
\label{eq:alpha_3D_simplified}
\end{equation}
By using Eq.(\ref{eq:alpha_3D_simplified}), on expresses
$\rho^{(3)}(\epsilon)$ as
\begin{equation}
\rho^{(3)}(\epsilon)=\frac{\rho_0^{(3)}}{\sqrt{2}}\frac{\sqrt{1+\alpha_R+\sqrt{(1+\alpha_R)^2+\alpha_I^2}}}{\sqrt{(1+\alpha_R)^2+\alpha_I^2}}\approx \frac{\rho_0^{(3)}}{\sqrt{1+\frac{3}{4\pi \epsilon_F^2 \tau_\pi^2}\bigl(1-\frac{\pi \sqrt{3 |\epsilon|\tau}}{2}\bigr)}}
\label{eq:rho_3D_simplified}
\end{equation}
where $\alpha_R(\epsilon)$ and $\alpha_I(\epsilon)$ are real and
imaginary parts of $\alpha_3(\epsilon)$, respectively.

Eq.(\ref{eq:rho_3D_simplified}) shows that a small dip of the DOS is
formed at the middle of a half--filled band of a 3D simple cubic
lattice due to substitutional impurities (see Fig.\ref{fig:rho_2D}b).
The depth of this dip increases with disorder strength as
\begin{equation}
\rho_0^{(3)}(0)=\frac{\rho_0^{(3)}}{\sqrt{1+\frac{3}{4\pi\epsilon_F^2 \tau_\pi^2}}}
\label{eq:rho_3D_dip}
\end{equation}
The results obtained for the DOS of 2D systems,
Eqs.(\ref{eq:rho_from_alpha})-(\ref{eq:rho_2D_singularity}), and 3D
systems,
Eqs.(\ref{eq:alpha_3D_simplified})-(\ref{eq:rho_3D_dip}), show
that in contrast to 1D systems substitutional impurities tend to
reduce the DOS on the Fermi surface of 2D and 3D crystals at half
filling. Indeed, the Dyson singularity, expressing the enhancement of
the states density in the 1D lattice with off--diagonal disorder, is
mediated by
impurities,\cite{dyson53a,weissman75a,gorkov76a,ovchinnikov77a,gogolin77a,gredeskul78a,hirsch76a,eggarter78a,inui94a}
since the DOS of a regular 1D lattice is a smooth function of the
energy of a half--filled band.

The mechanism of the relative reduction of the DOS in 2D
[Eq.(\ref{eq:rho_2D_sol})] and 3D [Eq.(\ref{eq:rho_3D_simplified})]
systems is Umklapp scattering, the same that increases the 1D DOS.
However, the nested Fermi surface of a 2D square crystal with
nearest--neighbor hopping contains also saddle points at $\{\pm
\frac{\pi}{a},0\}$ and $\{0,\pm\frac{\pi}{a}\}$ which cause the van
Hove singularity in the DOS. Umklapp scattering in this case weakens
the central van Hove peak in the DOS, nevertheless it could not damp
the peak or reverse its sign.
We understand the different effects of substitutional impurities on the DOS of 1D and higher dimensional systems in the following way: The main mechanism of localization in a 1D disordered system is backward scattering on impurities.\cite{lifshits88a,berezinskii73a,mott61a} In the process of scattering on an impurity, simultaneous Bragg reflection at half--filling reverses backward scattering to forward scattering and vice versa. Therefore for vanishingly small forward scattering amplitude\cite{gorkov76a,gogolin77a} Bragg reflection prevents localization due to impurity scattering. However, the situation is different for 2D and 3D systems. According to the intuitive discussion given by Bergmann,\cite{bergmann83a} localization in 2D systems is due to interference of the electronic wave, returning to the starting point after multiple scattering on impurities with only a small change of the momentum at each act of scattering, with the wave on the time--reversed path (see also[\onlinecite{altshuler85a}]). In this case, the Bragg reflection accompanying the impurity scattering can not destroy the picture of interference and as a result can not completely delocalize all states.

The addition of next--nearest--neighbor
hopping terms into the model with strength $t'$ splits saddle points
from the nested Fermi surface. For the energy spectrum
$\epsilon({\bf k}) = -t[\cos p_x a +\cos p_y a- \frac{t'}{t} \cos p_x a \cos
p_y a + \frac{\mu}{2}]$ the saddle points again lie at $\{\pm \frac{\pi}{a},0\}$ and
$\{0,\pm\frac{\pi}{a}\}$. However, the optimal nested Fermi surface is
realized at $t'/t=0.165$ and $\mu=-0.56$, with the new nesting vector ${\bf
  Q}_0^*=0.91 {\bf Q}_0$.\cite{ruvalds95a}  Our method can be applied also for this 
dispersion band. The singular
blocks are again calculated after separating the large momentum
transfer ${\bf Q}_0^*$ and linearizing the energy spectrum around the
Fermi surface. As a result the same expression
(\ref{eq:rho_2D_singularity})) for the 2D DOS is obtained. However,
$\rho_0^{(2)}$ in this case has no singularity on the Fermi surface at
half--filling. Therefore $\rho^{(2)}(\epsilon)$ decreases with the
energy around the Fermi surface for half filling and vanishes on it.

\section{Conductivity}
\label{sec:conductivity}

\begin{figure}
  \centerline{\epsfbox[148 446 512 669]{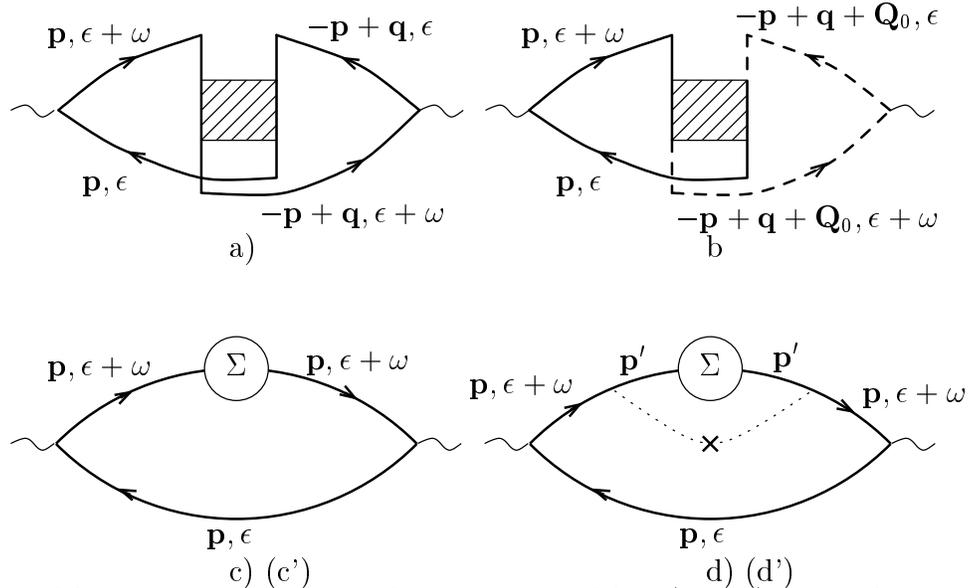}}
\caption{Diagrams giving the first quantum correction to the conductivity. a) and b) characterize the diffusion coefficient contributions to $\sigma(\omega)$ from normal and umklapp scattering channels, respectively. $\Sigma$ in diagrams c)-d), for the contributions from the DOS to $\sigma(\omega)$, is the sum of all self--energy contributions shown in Fig.\ref{fig:self_energy_diagrams}. Two other diagrams c') and d') differ from c) and d) through the direction of the electron lines.}
\label{fig:conductivity}
\end{figure}

The DC-conductivity for the 2D Anderson model at zero temperature has
been computed in [\onlinecite{stein80a}] and the behavior of $\sigma$
as a function of the Fermi energy and the disorder has been studied.
However, these numerical results provide limited insight into the
physical origin of the processes giving contributions to the
conductivity. In this section, we study low temperature quantum
contributions to the conductivity for the model under consideration.
Maximally crossed or ``fan'' diagrams are responsible for the quantum
interference corrections to the conductivity in the weak localization
theory.\cite{gorkov79a} Interference between electronic wave functions
in the process of multiple scattering on randomly distributed
impurities changes the mobility of an electronic system and ``fan''
diagrams give a contribution to the diffusion coefficient. The
conductivity of simple crystalline systems with substitutional
impurities can also be affected through changes in the DOS due to
Bragg reflection in the scattering processes on the impurities which
becomes essential for an electronic band close to half--filling. The
diagrams which describe first order weak localization corrections to
the conductivity and represent both the diffusion coefficient and the
DOS contributions, are drawn in Fig.\ref{fig:conductivity}.

Since interference effects are essential for low--dimensional systems,
we here will calculate the quantum corrections to the conductivity
only for a 2D square lattice with substitutional impurities. The
localization problem for a 1D lattice with off-diagonal impurities has
been calculated in [\onlinecite{weissman75a,gorkov76a,ovchinnikov77a,gogolin77a,gredeskul78a,hirsch76a,eggarter78a,inui94a}].

The quantum correction to the conductivity is calculated according to
the Kubo expression
\begin{equation}
\sigma_{\alpha,\beta}(\omega)=i\frac{Ne^2}{m\omega}\delta_{\alpha\beta}+
\frac{2e^2}{\omega}\int \frac{{\rm d}\epsilon}{2\pi} \int
\frac{{\rm d}^2p}{(2\pi)^2} v_\alpha({\bf p}) v_\beta({\bf p}) 
\langle G({\bf p},\epsilon+\omega) G({\bf p},\epsilon) \rangle
\label{eq:kubo}
\end{equation}

where $G({\bf p},\epsilon)$ is the Green's function and $v_\alpha({\bf
  p})=\frac{\partial\epsilon}{\partial p_\alpha}$ with $\alpha=(x,y)$
is a component of the electron velocity; the bracket $\langle \dots \rangle$ denotes averaging over the impurity realizations.

Far from the half--filled energy band, maximally crossed diagrams with
normal scattering are responsible for logarithmic corrections to the
conductivity of the 2D weakly disordered non--interacting electron
gas.\cite{gorkov79a} These diagrams can be redrawn as a ladder series in
the particle--particle channel as shown in
Fig.\ref{fig:conductivity}a.

Bragg reflection is intensified as half--filling is approached. In
this case, an act of electron scattering on an impurity may be
accompanied by Bragg reflection on the Brillouin zone boundary. The
new diagrams which give a contribution to the conductivity as
half--filling is approached are drawn in
Figs.\ref{fig:conductivity}b-d'.

The contribution of the diagrams in Fig.6a to the conductivity,
$\delta\sigma_a(\omega)$ can be expressed as follows according to
the Kubo expression Eq.(\ref{eq:kubo}):
\begin{equation}
\delta\sigma_a(\omega) = \frac{2e^2}{\omega}\int \frac{{\rm d}\epsilon}{2\pi} \int
\frac{{\rm d}^2p}{(2\pi)^2} C_N({\bf q},\epsilon,\omega) A_\alpha^N({\bf q},\epsilon,\omega)
\label{eq:delta_sigma}
\end{equation}
where $C_N(({\bf q},\epsilon,\omega)$ is the cooperon block
Eq.(\ref{eq:Cn}) for normal scattering and
\begin{equation}
A_\alpha^N({\bf q},\epsilon,\omega)=\int \frac{{\rm d}^2p}{(2\pi)^2}
v_\alpha({\bf p})v_\alpha(-{\bf p}+{\bf q})G({\bf p},\epsilon+\omega)G(-{\bf p}+{\bf q},\epsilon+\omega)G({\bf p},\epsilon)G(-{\bf p}+{\bf q},\epsilon)
\label{eq:A_alpha}
\end{equation}

For a tight--binding model with nearest--neighbor hopping, the
$\alpha=(x,y)$ component of the velocity $v_\alpha({\bf p})$ is given
as $v_\alpha({\bf p})=ta \sin(p_\alpha a)$. The integration over ${\bf
  p}$ can be written as $\int\frac{{\rm d}{\bf S}}{(2\pi)^2 |{\bf
    v}_p|}\int {\rm d} \xi$, and then the integral over the energy
variable $\xi$ is done easily. The van Hove singularity, arising at
saddle points of the Fermi surface for the half--filled band when
$|{\bf v}_{{\bf p}}|=t a \sqrt{\sin^2 p_x a + \sin^2 p_y a}$ vanishes,
is removed due to the $v_\alpha^2({\bf p})$ term under the integral
over the Fermi surface ${\rm d}{\bf S}$. Therefore the value of
$A_\alpha(q,\epsilon,\omega)$ does not strongly differ from that for
the free--electron gas model and we get
\begin{equation}
A_\alpha({\bf q},\epsilon,\omega)=-p_F\tau^2l\Theta(-\epsilon(\epsilon+\omega))
\label{eq:A_alpha_sol}
\end{equation}
Using Eqs.(\ref{eq:Cn}) and (\ref{eq:A_alpha_sol}) for $C_N({\bf
  q},\epsilon,\omega)$ and $A_\alpha({\bf q},\epsilon,\omega)$ in
Eq.(\ref{eq:delta_sigma}) one gets
\begin{equation}
\delta\sigma_a(\omega) = - \frac{e^2\tau}{2\pi^2\tau_N}\ln\Bigl(\frac{1}{\frac{\tau}{\tau_\pi}-i|\omega|\tau|}\Bigr)
\label{eq:delta_sigma_sol}
\end{equation}

Far from half--filling and far from other commensurate points the
Bragg reflection is weakened and $\tau \approx\tau_N\ll \tau_\pi$.
Therefore $\delta\sigma_a(\omega)$ in this case represents the
conventional weak localization correction to the
conductivity.\cite{gorkov79a} $\delta\sigma_a(\omega)$ is reduced
close to half--filling when umklapp scatterings are enhanced and
$\tau\approx\tau_\pi \ll \tau_N$.

The diagrams which give an essential contribution to the conductivity
at half--filling are shown in
Fig.\ref{fig:conductivity}b-\ref{fig:conductivity}d'.  The appearance
of these new diagrams is due to intensified scattering in the
$\pi$--channel with large momentum transfer. The main quantum
correction to the diffusion coefficient at half--filling comes from
the ``fan'' diagrams in Fig.\ref {fig:conductivity}b with the
$\pi$--cooperon block. The expression corresponding to this diagram is
\begin{equation}
\delta\sigma_b(\omega)=\frac{2e^2}{\omega}\int \frac{{\rm d}\epsilon}{2\pi} \int
\frac{{\rm d}^2q}{(2\pi)^2} C_\pi({\bf q},\epsilon,\omega)A_\pi({\bf q},\epsilon,\omega)
\label{eq:delta_sigma_b}
\end{equation}
where $C_\pi({\bf q},\epsilon,\omega)$ is the $\pi$--cooperon given by
Eq.(\ref{eq:Cp_sol}) and
\begin{equation}
\begin{split}
  A_\pi({\bf q},\epsilon,\omega)=&\int \frac{{\rm d}^2p}{(2\pi)^2}
  v_\alpha({\bf p})v_\alpha(-{\bf p}+{\bf q}+{\bf Q}_0)
  G({\bf p},\epsilon+\omega)G({\bf p},\epsilon)\\
  &G(-{\bf p}+{\bf q}+{\bf Q}_0,\epsilon)G(-{\bf p}+{\bf q}+{\bf
    Q}_0,\epsilon+\omega)
\end{split}
\label{eq:A_pi}
\end{equation}
The calculation of Eq.(\ref{eq:A_pi}) is similar to that for $A_N({\bf
  q},\epsilon,\omega)$. Taking into account the condition that
$v_\alpha({\bf k}+{\bf Q}_0)=-v_\alpha({\bf k})$ we obtain the result
for $A_\pi({\bf q},\epsilon,\omega)$
\begin{equation}
A_\pi({\bf q},\epsilon,\omega) = \frac{p_F l \tau^2}{(1-2i\tau|\epsilon|)(1-2i\tau|\epsilon+\omega|)}\Bigl\{\frac{1}{1-i\tau|2\epsilon+\omega|}\Theta(\epsilon(\epsilon+\omega))+\Theta(-\epsilon(\epsilon+\omega))\Bigr\}
\label{eq:A_pi_sol}
\end{equation}
Substituting Eqs.(\ref{eq:Cp_sol}) and (\ref{eq:A_pi_sol}) in
Eq.(\ref{eq:delta_sigma_b}), the expression for
$\delta\sigma_b(\omega)$ can be reduced by some simple calculations to
\begin{equation}
\delta\sigma_b(\omega)=\frac{e^2\tau}{2\pi^2\omega\tau_\pi}\Bigl\{\bigl(\frac{\tau}{\tau_\pi}\bigr)^2I(\frac{\tau}{\tau_\pi},\omega \tau)+\frac{\omega}{2}\Bigr\}
\label{eq:delta_sigma_b_sol}
\end{equation}
with
\begin{equation}
\begin{split}
  I&(\frac{\tau}{\tau_\pi},\overline{\omega}=\omega
  \tau)=\frac{i}{2\tau}\int_0^1{\rm d}x
  \int_0^{\infty}{\rm d}z \frac{1}{(z+i)^3}
\frac{1}{(z+\overline{\omega}+i)^2-x+(\frac{\tau}{\tau_\pi})^2}\\
  =&\frac{i}{4\tau}\Bigl\{2i\overline{\omega}\bigl[
\frac{1}{\overline{\omega}^2-1+(\frac{\tau}{\tau_\pi})^2}
-\frac{1}{\overline{\omega}^2+(\frac{\tau}{\tau_\pi})^2}\bigr]+
  \bigl[1+\frac{1}{(\overline{\omega}-\sqrt{1-(
\frac{\tau}{\tau_\pi})^2})^2}\bigr] \ln\bigl(1-i\overline{\omega}
+i\sqrt{1-(\frac{\tau}{\tau_\pi})^2}\bigr)\\
  &+\bigl[1+\frac{1}{(\overline{\omega}+\sqrt{1
-(\frac{\tau}{\tau_\pi})^2})^2}\bigr]\ln\bigl(1-i\overline{\omega}
-i\sqrt{1-(\frac{\tau}{\tau_\pi})^2}\bigr)+
  \bigl[\frac{1}{(\frac{\tau}{\tau_\pi}-i\overline{\omega})^2}
-1\bigr]\ln\bigl(1-i\overline{\omega}+\frac{\tau}{\tau_\pi}\bigr)\\
  &+\bigl[\frac{1}{(\frac{\tau}{\tau_\pi}+i\overline{\omega})^2}
-1\bigr]\ln\bigl(1-i\overline{\omega}-\frac{\tau}{\tau_\pi}\bigr)\Bigr\}
\label{eq:I}
\end{split}
\end{equation}
The expression for $\delta\sigma_b(\omega)$ can be presented for two
limiting cases, namely near the middle of the band when
$\tau\approx\tau_\pi \ll \tau_N$ and far from half--filling when
$\tau\approx\tau_N\ll \tau_\pi$ by using
Eq.(\ref{eq:I}) in (\ref{eq:delta_sigma_b_sol}):
\begin{equation}
\delta\sigma_b(\omega)=
\begin{cases}
  \frac{e^2}{4\pi^2}\ln(1-\frac{\tau}{\tau_\pi}-i\omega\tau)+i\frac{e^2}{8\pi^2\omega\tau_\pi}& \mbox{for}\quad\frac{\tau}{\tau_\pi}\to 1\\
  {\rm
    const}+i\frac{e^2}{4\pi^2\omega\tau_\pi}\bigl(\frac{\tau}{\tau_\pi}\bigr)^2\ln2&
  \mbox{for}\quad\frac{\tau}{\tau_\pi}\approx\frac{\tau_N}{\tau_\pi}\to
  0
\end{cases}
\label{eq:delta_sigma_b_approx}
\end{equation}
So the quantum correction to the conductivity corresponding to the diagram in Fig.\ref{fig:conductivity}b decreases $\sigma(\omega)$ logarithmically with the
external frequency near half--filling. The contribution of
Fig.\ref{fig:conductivity}b vanishes through a small offset from the
middle of the band.

Analyzing the effect of substitutional impurities on the conductivity
by means of the diffusion coefficient, which is expressed by the
diagrams a) and b) in Fig.\ref{fig:conductivity}, it can be seen that
the logarithmic correction to the localization correction to $\sigma(\omega)$ is preserved irrespectively of the band filling. However, the
coefficient of the logarithm of the quantum correction to the
conductivity is changed from $\frac{e^2}{2\pi^2}$ in Eq.(\ref{eq:delta_sigma_sol}) far from
half--filling to $\frac{e^2}{4\pi^2}$ in Eq.(\ref{eq:delta_sigma_b_approx}) close to
half--filling. On the other hand, the $\pi$--scattering mechanism
gives a perturbative contribution to the dielectric constant
$\epsilon'(\omega)\sim {\rm Im}\sigma(\omega)$. The
contribution to $\epsilon'(\omega)$ increases close to half--filling.
This also means that there exist a few delocalized states at the center of
the band.

The diagrams in Fig.\ref{fig:conductivity}c-d' represent the
quantum correction to the conductivity due to changes in the DOS. The
expression corresponding to the sum of these diagrams can be presented
as
\begin{equation}
2(\delta\sigma_c(\omega)+\delta\sigma_d(\omega))=\frac{2e^2}{\omega}
\int \frac{{\rm d}\epsilon}{2\pi} \int \frac{{\rm d}^2q}{(2\pi)^2}
\alpha({\bf q},\epsilon+\omega)B({\bf q},\epsilon,\omega)
\label{eq:delta_sigma_cd}
\end{equation}
where
\begin{equation}
\alpha({\bf q},\epsilon)=C_\pi({\bf q},\epsilon)-\frac{2 \tau}{\tau_\pi}(1-\frac{\tau}{\tau_\pi})D_\pi({\bf q},\epsilon)
\label{eq:alpha2}
\end{equation}
and
\begin{equation}
B({\bf q},\epsilon,\omega)=\frac{p_Fl\tau^2}{(1-2i\tau|\epsilon+\omega|)^2}\Bigl\{(\frac{\tau}{\tau_N}-2)\Theta(-\epsilon(\epsilon+\omega))+\frac{1}{1-i\tau|\epsilon+\omega|-i\tau|\epsilon|} \Theta(\epsilon(\epsilon+\omega))\Bigr\}
\label{eq:B}
\end{equation}
The fact that the contributions from diagrams c' and d' in
Fig.\ref{fig:conductivity} are equal to those from c and d is taken
into account in Eq.(\ref{eq:delta_sigma_cd}). The cooperon and diffusion insertions into the
Green's function are given by Eqs.(\ref{eq:Cp0}) and (\ref{eq:Dp0}), respectively.  After
some routine calculations, Eq.(\ref{eq:delta_sigma_cd}) for the
correction to the conductivity due to the DOS is reduced to the form
\begin{equation}
\begin{split}
  2(\delta\sigma_c(\omega)+&\delta\sigma_d(\omega))=
  \frac{e^2\tau^3}{2\pi^2\omega\tau_\pi^3}
  \bigl[1-\frac{2 \tau}{\tau_\pi}(1-\frac{\tau}{\tau_\pi})\bigr] \Bigl\{
  I(\frac{\tau}{\tau_\pi},\overline{\omega}=\omega\tau)\\
  & +\frac{\omega}{2} (\frac{\tau}{\tau_N}-3) \ln
  \frac{1-(\frac{\tau}{\tau_\pi}))^2+(1-2i\omega\tau)^2}{(1-2i\omega\tau)^2+(\frac{\tau}{\tau_\pi})^2}
  \Bigr\}
\end{split}
\label{eq:delta_sigma_cd_simplified}
\end{equation}
Where $I(\frac{\tau}{\tau_\pi},\overline{\omega}=\omega\tau)$ is given by Eq.(\ref{eq:I}).
These contributions vanish as expected far from half--filled band,
when $\tau_\pi \gg \tau\approx\tau_N$. However, they give a large
contribution to the conductivity near half--filling,
$\tau\approx\tau_\pi \ll \tau_N$, giving rise to a rapid increasing of
the conductivity with the external frequency. So, the total
contribution to the conductivity is obtained by summing Eqs.(\ref{eq:delta_sigma_sol}),(\ref{eq:delta_sigma_b_approx}),
and (\ref{eq:delta_sigma_cd_simplified}). Far from half--filling, only the normal cooperon gives a
contribution and, as a result $\delta\sigma_a(\omega)$ survives:
\begin{equation}
\delta\sigma(\omega)=-\frac{e^2}{2\pi^2}\ln(-\frac{1}{i|\omega|\tau})
\label{eq:delta_sigma_away_from_half_filling}
\end{equation}
The quantum correction to the conductivity near half--filling is due
to the $\pi$--cooperon and
\begin{equation}
\delta\sigma(\omega)=-\frac{5 e^2}{2\pi^2}\ln(-\frac{1}{i|\omega|\tau_\pi})
\label{eq:delta_sigma_at_half_filling}
\end{equation}
with $\tau \to \tau_\pi$. Although the conductivity decreases with the
external frequency, the diffusion coefficient increases with
approaching half--filling for given $\omega$, which means a partial
lifting of localization at the center of the band due
to a few delocalized states. The rapid decrease
of the conductivity (Eq.(\ref{eq:delta_sigma_at_half_filling})) near
half--filling is the result of the impurity effect on the DOS.

\section{Conclusion}
\label{sec:conclusion}

The character of localization in the 2D disordered electronic gas has
been a subject of debate for a long time. The question whether
delocalized states exist in the center of the band of 2D disordered
systems or not is also one of the crucial points for the integer
quantum Hall effect.

Searches for delocalized states have been mainly concentrated around
the 2D Anderson model. Expectations are connected with
the existence of the van Hove singularity in the band center of the
square lattice, which might give rise to a delocalization of states at
half--filling. Various numerical approaches have been applied to
compute the DOS,\cite{hu84a,schreiber98a} the localization
length,\cite{soukoulis82a,kramer83a,schreiber98a,schreiber98b,kramer93}
and the conductivity\cite{stein80a,lee81a} of the 2D Anderson model
with diagonal and off--diagonal disorder. The computations show that
although even a small concentration of diagonal disorder suppresses
the van Hove singularity, off--diagonal disorder preserves it,
however with a modified shape.

The exponential localization of all states for diagonal
disorder\cite{lee81a,kramer83a} has been revealed by computational approaches
whereas the existence of quasi--localized states has been predicted
for the 2D Anderson model with off--diagonal
disorder.\cite{soukoulis82a,schreiber98a,schreiber98b} In other words, the computation of the effects of off--diagonal disorder provides evidence for the transformation from exponential to power--like localization as half--filling is approached.

In this paper we tried to give a complete picture for the DOS and the conductivity, and also to give some physical insight into the processes occuring near half--filling in simple lattices with substitutional impurities.

The impurity concentration is considered to be small, so that the condition $\epsilon_F \tau \gg 1$ or $p_F l \gg 1$ is satisfied and the weak localization approximation is applicable.

Apart from normal scattering with small momentum transfer, umklapp scattering with large momentum transfer is shown to be essential near half--filling. This new scattering involves coherent reflection of an electron on the boundary of the Brillouin zone for each act of scattering.

Our diagrammatical approach shows that the new singular blocks, namely the $\pi$--cooperon $C_\pi({\bf q},\epsilon,\omega)$ and the $\pi$--diffuson $D_\pi({\bf k},\epsilon,\omega)$, give a large contribution to the DOS and the conductivity at half--filling.
The dependence of the DOS on the energy and the impurity strength near the middle of the band have been determined by summing the leading logarithmically divergent contributions. The results obtained [Eqs.(\ref{eq:rho_2D_sol}) and (\ref{eq:rho_3D_simplified})] for the DOS show that weak disorder due to substitutional impurities does not remove the van Hove singularity at the center of the 2D band. However its energy dependence is strongly changed. The impurity effect on the 3D DOS is a shallow dip on the smooth background of the bare DOS in the middle of the band.

The effect might be observable in the temperature dependence of the magnetic susceptibility according to
\begin{equation}
\chi_{(d)}(T)=2\mu_B^2\int \frac{{\rm d}\epsilon}{4 T} \cosh^{-2}(\frac{\epsilon}{2T})\rho^{(d)}(\epsilon)
\label{eq:susceptibility}
\end{equation}
Where $\mu_B$ is the Bohr magneton.
Straightforward calculation using Eq.(\ref{eq:rho_2D_sol}) for the 2D DOS gives

\begin{displaymath}
\chi_{2D}(T)\propto \ln^{1/2}(\frac{1}{T \tau_\pi})
\end{displaymath}

For $d=3$, the correction to the magnetic susceptibility can again be calculated according to Eqs.(\ref{eq:susceptibility}) and (\ref{eq:rho_3D_simplified}). However, this correction is negligibly small.

The first logarithmic corrections to the conductivity of the 2D square lattice with substitutional impurities come from both the diffusion coefficient and the DOS. The corrections due to the diffusion coefficient, $\delta\sigma_a(\omega)+\delta\sigma_b(\omega)$, given by Eqs.(\ref{eq:delta_sigma_sol}) and (\ref{eq:delta_sigma_b_approx}), decrease in magnitude as half--filling is approached while they remain logarithmically dependent on the external frequency. Such a partial lifting of localization may be due to a few delocalized states in the center of the band.\cite{soukoulis82a,kramer83a,schreiber98a,schreiber98b,kramer93} The imaginary part of $\sigma(\omega)$, which corresponds to the dielectric constant, increases close to half--filling. This fact supports the picture of an increase of the electronic mobility at the center of the band.

Nevertheless, the contribution to $\sigma(\omega)$ due to the DOS suppresses the relative increase in the conductivity coming from the diffusion coefficient.



\end{document}